\documentstyle[PASJadd]{PASJ95}
\input epsf.sty

\newcommand{\vol}{52}
\newcommand{\no}{1}
\newcommand{\lett}{}
\newcommand{\spage}{1}
\newcommand{\rdate}{1999 October 28}
\newcommand{\adate}{1999 December 21}
\begin{document}
\title{Infrared Imaging of $z=$ 2.43 Radio Galaxy B3~0731+438 with
the Subaru Telescope
--- Detection of H$\alpha$ Ionization Cones of a Powerful Radio Galaxy
}

\author{
Kentaro {\sc Motohara},
Fumihide {\sc Iwamuro},
Hiroshi {\sc Terada},
Miwa {\sc Goto},\\
Jun'ichi {\sc Iwai},
Hirohisa {\sc Tanabe},
Tomoyuki {\sc Taguchi},
Ryuji {\sc Hata},\\
Toshinori {\sc Maihara},\\
{\it Department of Physics, Kyoto University, Kitashirakawa, Kyoto 606-8502}\\
{\it E-mail(KM): motohara@cr.scphys.kyoto-u.ac.jp}\\
Shin {\sc Oya},\\
{\it Communications Research Laboratory, Koganei, Tokyo 184-8975}\\
Masanori {\sc Iye},\\
{\it Optical and Infrared Astronomy Division, National Astronomical Observatory, Mitaka, Tokyo 181-8588}\\
George {\sc Kosugi},
Jun'ichi {\sc Noumaru},\\
Ryusuke {\sc Ogasawara},
Toshinori {\sc Sasaki},
and Tadafumi {\sc Takata}\\
{\it Subaru Telescope, National Astronomical Observatory, 650 North Aohoku Place, Hilo, HI 96720, USA}\\
}
\abst{
We report on infrared imaging observations of the $z=2.429$ radio galaxy B3
 0731+438 with the Subaru telescope. The images were taken with the
 $K^{\prime}$-band filter and the 2.25 $\mu$m narrow-band filter to
 examine the structure and properties of the H$\alpha$+[N
 {\sc ii}] $\lambda\lambda$ 6548, 6583 emission-line components. 
 The H$\alpha$+[N {\sc ii}] emission-line image shows biconical lobes with an
 extent of 40 kpc, which are aligned with the radio axis.
The rest-frame equivalent widths of the emission lines at these cones are as
 large as 1100 $\rm \AA$, and can be well explained by a gas-cloud model
 photoionized by power-law continuum radiation.
The isotropic ionizing photon luminosity  necessary to ionize
 the hydrogen gas in these cones amounts to $10^{57}$(photons s$^{-1}$),
 which is larger than that in the majority of radio-loud QSOs.
From these results, we propose that the H$\alpha$ alignment effect
 in this object is produced by biconical gas
 clouds, which are swept up by the passage of radio jets, 
 and are ionized by strong UV radiation from a hidden AGN.
The continuum image consists of two components, a
 stellar-like point source and an extended diffuse galaxy.
These are supposed to be a type-2 AGN and its host galaxy.
The SED is fitted by a combination of spectra of a reddened dust-scattered
 AGN and an instantaneous starburst population of 500 Myr old.
The stellar mass of the galaxy is estimated to be $3\times10^{11}~M_{\odot}$
, which is as large as that of typical 3C radio galaxies at $z=1$.
}

\kword{AGN: ionization cone --- galaxies: active --- galaxies:
individual (B3~0731+438) ---  infrared: galaxies }

\maketitle
\thispagestyle{headings}

\section
{Introduction}

It is well known that in high-redshift radio galaxies, the elongated
morphologies 
of the emission lines and rest-frame UV continuum light are often
aligned with the radio axes (McCarthy et al.\ 1987; Chambers et al.\ 1987).
The cause of this ``alignment-effect'' is still unclear, although it is almost
certain that this phenomenon is due to the anisotropic structures of AGNs at
the centers of the galaxies.  
Several hypotheses have been suggested, such as induced star
formation by the passage of radio jets (Rees 1989; Begelman, Cioffi 1989;
de Young 1989), inverse-Compton scattering of cosmic-background
radiation (Daly 1992a, 1992b), dust or electron scattering
of anisotropic radiation from a hidden AGN
(di Serego Alighieri et al.\ 1989; Fabian 1989), and nebular emission from
surrounding clouds ionized by UV radiation from a central engine (Dickson
et al.\ 1995). 

To examine the structure and properties of the H$\alpha$ emission-line
region in a high-redshift radio galaxy, we carried out
high-resolution infrared imaging observations of a powerful radio
galaxy, B3~0731+438, at $z=$ 2.429 in the $K^{\prime}$-band and the 2.25 
$\mu \rm m$ narrow-band. 
This object is a typical FR II radio galaxy with strong double-lobed
radio-emitting hot spots and a central core (Carilli et al.\ 1997).
The optical $R$-band image shows a diffuse morphology with a few knots,
while the narrow-band image presents aligned Ly$\alpha$ emission clouds
(McCarthy 1991). The optical spectrum displays an extremely strong L$\alpha$
emission line with a rest-frame equivalent width of 900 $\rm \AA$
(McCarthy 1991), which is one of the the largest values among known high-$z$
radio galaxies. 
Infrared spectra reveal a strong H$\alpha$+[N {\sc ii}] emission line
(McCarthy et al.\ 1992; Eales, Rawlings 1993; Evans 1998),
contributing  $20$ -- $30 \%$ flux in the $K$-band. Diagnostic emission-line
ratios are consistent with the presence of a Seyfert 2 nucleus (Evans
1998).

We describe our observations and data reductions in section 2, and
present the results in section 3.
Discussions are presented in section 4 and we summarize them in section 5.
Cosmological constants are assumed to be $H_0=50$ km s$^{-1}$ Mpc$^{-1}$ and
$q_0=0.1$ throughout this paper.
The scale at $z$ = 2.429 is thus 11.3 kpc/$^{\prime\prime}$, and the
look-back time is 13 Gyr, while the age of the universe is 16.5 Gyr.

\section
{Observations and Data Reduction}
\subsection{Observations}
B3~0731+438 was observed using a 2.25 $\mu$m narrow-band filter
($\lambda/\Delta\lambda=100$) on 1999
February 25 and 27, and using a $K^{\prime}$-band filter on February
27, with a near-infrared camera CISCO (Motohara et al.\ 1998) mounted
on the Cassegrain focus of the Subaru 
telescope. The total field of view of the camera was $\sim
2^{\prime}\times 2^{\prime}$ with a pixel scale of
$0.\!\!^{\prime\prime} 116$ pixel$^{-1}$. 
The exposure time of a single frame was 60 s for the narrow-band
and 20 s for the $K^{\prime}$-band images.
To subtract the background sky emission, we nodded the telescope
slightly($\sim10^{\prime\prime}$) and acquired 
6 (narrow-band) or 12 ($K^{\prime}$-band) frames at each of eight different
positions, resulting in 48 or 96 frames
per each dataset, respectively. The total exposure time was 5760 s for
the narrow-band and 1920 s for the $K^{\prime}$-band.
The atmospheric conditions were photometric and the seeing was
stable during the observation, which was 0$.\!\!^{\prime\prime}$6 on
February 25 and 0$.\!\!^{\prime\prime}$4 on 27. The logs of the
observations are summarized in table 1. 

\subsection{Data Reduction}
We carried out data reduction of the $K^{\prime}$-band frames as follows.
First, a ``standard $K^{\prime}$ flat'' frame was produced by
median-averaging 250 available $K^{\prime}$-band frames taken on
February 23, 25, and 27, excluding the B3~0731+438 frames.
Then, every B3~0731+438 frame was divided by the standard $K^{\prime}$
flat frame.
After applying cosmetic corrections for bad pixels, we made a ``sky''
frame by median-averaging the frames of which prominent objects were masked
and replaced by the surrounding sky value with appropriate noise.
We smoothed this sky frame with a 20$\times$20 pixels boxcar
filter, normalized the average of the pixel value to unity and
multiplied it by the standard $K^{\prime}$ flat frame to make a
``self-calibrated $K^{\prime}$ flat'' frame. 

Using this self-calibrated  $K^{\prime}$ flat frame, a sky frame was
generated again by following the above procedure
from the beginning, and was subtracted from each frame.
The 12 frames in each set were shifted by a sub-pixel offset according to 
a reference star in the frames and averaged to create a ``minor'' frame.
The final $K^{\prime}$-band image was produced by median-averaging these
minor frames. The seeing size of the final image was
0$.\!\!^{\prime\prime}$4. 

Narrow-band frames were processed in the same way as for the
$K^{\prime}$-band, except that no self-calibrated flat frame was
produced, due to the small sky background flux in this band. 
Two versions of final narrow-band images were made: one was a ``total''
image produced from all the frames; the other was a ``good-seeing''
image produced only from the frames taken on February 27.  
The seeing size was 0$.\!\!^{\prime\prime}$6 for the total image and
0$.\!\!^{\prime\prime}$4 for the good-seeing image.
We created a color image of B3~0731+438 from the $K^{\prime}$-band and
the ``good-seeing'' narrow-band image,
which is shown in figure 1.

To investigate the emission-line and continuum properties of B3
0731+438, we made two post-processed images.
One was a line-free continuum image ($K$-continuum image) produced by
subtracting the scaled narrow-band image from the $K^{\prime}$-band image.
The other was a continuum-subtracted emission-line image
(H$\alpha$+[N {\sc ii}] image) produced by subtracting the scaled $K$-continuum
image from the total narrow-band image. 
We present these images in figure 2 together with their contours.

\section{Results}
\subsection{Photometry}
The results of aperture photometry of B3~0731+438 are given in table 2
together with that of other observations. The aperture is
$9.\!\!^{\prime\prime}4$  diameter. 
The flux calibration was done using images of UKIRT faint standard
stars (FS 15, FS 21, and FS 23) taken before or after the observations.
The $K^{\prime}$- and narrow-band flux of the standard stars were
calculated by interpolating their $K$- and $H$-band flux.

\subsection{The $K$-Continuum Image}
The $K$-continuum image of  B3~0731+438 in figure
2b appears to comprise a very compact core
 and an extended diffuse nebula,
which can be seen in the deconvolved image (figure 2c) more clearly.
The compact core couldn't be resolved even with a spatial resolution as
high as 0$.\!\!^{\prime\prime}$4.
Therefore, we modeled the $K$-continuum image with a two-component
profile consisting of a stellar core and an exponential disk.

The profile of the core is assumed to be that of a field star fitted by a
modified moffat function.
The moffat function is written as
\begin{equation}
 I(x,y) = I_c\left[1+\left(\frac{r}{\alpha}\right)^{-\beta}\right],
\end{equation}
where $r$ is the radial distance from the center.
However, the stellar image is distorted elliptically because of
imperfect operation of the telescope (mirror support, auto-guiding
and so on) during the observations. We 
therefore had to introduce a PSF modification, defined as 
$r^2=X^2+Y^2$ and 
\begin{equation}
\scriptsize
\left(
\begin{array}{c}
X\\ Y
\end{array}
\right)
=
\left(
\begin{array}{cc}
\sqrt{1+e} & 0\\
0 & \sqrt{1-e}
\end{array}
\right)
\left(
\begin{array}{cc}
\cos{\Theta} & -\sin{\Theta}\\
\sin{\Theta} & \cos{\Theta}
\end{array}
\right)
\left(
\begin{array}{c}
x\\ y
\end{array}
\right),
\end{equation}
where $\Theta$ and $e$ are the position angle and the ellipticity of a
stellar image, respectively.
The values of these parameters for the field stars are
$\alpha=$ 0$.\!\!^{\prime\prime}$42, $\beta=3.1$, $\Theta =45^{\circ}$,
and $e=0.12$. 

We set five free parameters to reconstruct the $K$-continuum profile of
B3~0731+438. They are the peak height, the position relative to the core
profile, and the effective radius of the exponential disk and the peak height
of the core.  
The results of the fitting are shown in figure 3, and the obtained
 photometric data are listed in table 2. 
The FWHM of the exponential disk profile is 1$.\!\!^{\prime\prime}$6,
 which corresponds to 18 kpc. The peak of the disk is located at
 0$.\!\!^{\prime\prime}$24 south of the core.

\subsection{The Emission-Line Image}
The H$\alpha$+[N {\sc ii}] image in figure 2a shows a 
unique morphology.
Diffuse line emission extends out to 3$.\!\!^{\prime\prime}$3 from the
center, corresponding to 37 kpc.
They are aligned with the axis of the radio hot spots,
and both ends of the clouds fork into two directions.
Such a morphology suggests the existence of biconical clouds radiating
H$\alpha$+[N {\sc ii}] emission lines. 
The total flux of H$\alpha$+[N {\sc ii}] emission is $3.5\times10^{-18}$ W
m$^{-2}$, which corresponds to a luminosity of $3.2\times10^{37}$ W.

We used square aperture photometry
of the northern cone, the southern cone and the central core, which are 
marked as square boxes in figure 2.
The results are listed in table 3.
The most striking result is the extremely large rest-frame equivalent
width of more than 1000 $\rm \AA$ of the H$\alpha$+[N {\sc ii}] line in
the northern and
southern cones.
When the results of both cones are combined, the equivalent width at the
cones is $1098^{+516}_{-288}~\rm \AA$.

\section{Discussion}
Few narrow-band imagings aimed at the rest-frame optical emission-line of
powerful radio galaxies at a redshift of $z>2$ have been carried out,
because the lines redshift into the infrared wavelength. 
Armus et al. (1998) imaged the [O {\sc iii}] emission-line morphology
of $z=3.594$ radio galaxy 4C 19.71 through a 2.3 $\mu \rm m$ narrow-band
filter. They found a large ($\sim 70$ kpc) aligned nebular, whose
length is the same as that of the separation of the radio hot spots.
However, its morphology is like a ``corridor'' between the core and the
radio hot spots, and does not show a conic structure.
The total [O {\sc iii}] luminosity is $2\times 10^{37}$ W, which is
comparable to a H$\alpha$+[N {\sc ii}] luminosity of $3\times10^{37}$ W
of B3~0731+438, if we assume [O {\sc iii}]/H$\alpha$+[N {\sc ii}] = 1.
Egami et al. (1999) observed the H$\alpha$+[N {\sc ii}] emission-line
morphology of $z=2.269$ 4C 40.36, and found aligned H$\alpha$+[N {\sc
ii}] knots extended linearly over $\sim$ 20kpc, but no 
emission-line cone was seen.
They also found an unresolved ($<2$ kpc) continuum core, as we found
in B3~0731+438.

\subsection{Properties of Extended Emission Line Clouds}
Concerning the emission mechanism of the aligned
H$\alpha$+[N {\sc ii}] morphology of B3~0731+438, there are four major
hypotheses, as described before.
Among them, it is impossible to explain the observed emission-line
spectrum by the inverse-Compton scattering of the cosmic-background
radiation, from which no spectrum feature is expected.  

Regarding the second possibility of induced star formation, 
it is known that the Wolf--Rayet galaxy 
NGC 4861 shows a H$\alpha$+[N {\sc ii}] equivalent width larger than
900 $\rm \AA$ (McQuade et al.\ 1995).
Model spectra of a few-Myr old galaxies also 
show equivalent widths larger than 1000 $\rm \AA$ (Calzetti 1997).
On the other hand, the shape of the L$\alpha$ cloud with an equivalent
width of 900 $\rm \AA$ reported by McCarthy (1991), is pinched at the peak of 
the $K$-continuum image, and matches the H$\alpha$+[N {\sc ii}]
morphology that we have observed. Such a morphology suggests that
L$\alpha$ photons are radiated from the same region of the H$\alpha$+[N
{\sc ii}] cones, and that both emission lines have a common ionization
source. Since a stellar system cannot account for an
equivalent width of L$\alpha$ as large as 900 $\rm \AA$ (Charlot, Fall 1993), 
we infer that the alignment of the H$\alpha$+[N {\sc ii}] image 
is not caused by a star-forming region.

We cannot rule out the third possibility of scattered light of anisotropic
radiation from the central engine by dust or electrons, 
because the maximum H$\alpha$+[N {\sc ii}] equivalent widths of
high-redshift QSOs observed are 800 $\rm \AA$ (Baker et al.\ 1999; Hill et
al.\ 1993; Espey et al.\ 1989) and our 1$\sigma$ lower limit
is 800 $\rm \AA$.
We therefore evaluated the L$\alpha$/H$\alpha$ ratio to examine the
contribution of nebular emission to the H$\alpha$ flux,
because the extended L$\alpha$ emission is radiated as nebular
emission.
The observed L$\alpha$/H$\alpha$ is 5.5, assuming that all of the
L$\alpha$ flux ($3.1\times 10^{-18}~\rm W~ m^{-2}$; McCarthy 1991) is
radiated from the cones and that H$\alpha$/[N {\sc ii}]=4. 
On the other hand, L$\alpha$/H$\alpha$ under the case-B condition of
the low-density limit is 8.75 (Binette et al. 1992), and only a small
amount of dust, such as $E(B-V)=0.04$, halves the ratio.
We infer from this result that the majority of H$\alpha$ luminosity is
not scattered light from the central engine, but nebular emission
radiated from the cones, itself. 

Accordingly, we examined whether a gas cloud ionized by anisotropic UV
radiation from a hidden AGN can reproduce an equivalent width as large as
1100 $\rm \AA$. 
First, we extracted the physical properties of the cones and the
ionizing source from the data in table 3, using the
same method as that carried out by Baum and Heckman (1989).
Assuming the gas to be fully ionized and in the case-B condition, 
the luminosity of the H$\alpha$ emission line is written as 
\begin{equation}
 L({\rm H}\alpha)=n_{\rm e}^2 \alpha_{\rm H \alpha}^{\rm eff}h\nu_{\rm H \alpha} V f,
\end{equation}
where $n_{\rm e}$ is the electron density, $h$ the Planck constant, $\nu_{\rm
H\alpha}$ the frequency of the H$\alpha$ line, $V$ the volume of the line emitting
region, $f$ the volume filling factor, and
$\alpha_{\rm H\beta}^{\rm eff}=6.04\times10^{-14}$ (cm$^3$ s$^{-1}$)
(Osterbrock 1989), the H$\alpha$ recombination coefficient under
case-B. We assumed H$\alpha$/[N {\sc ii}] = 4
according to simulations by CLOUDY90 (see following)
and took $V$ for a cylinder and a half-cut
cylinder for the northern and southern cones, respectively.

Because we do not know the volume filling factor, $f$,
we must estimate it from direct measurements of 
other radio galaxies.
From measurements of the density of ionized gas using sulphur lines
in low-redshift radio galaxies,
their filling factor is estimated to be in range $10^{-4}$ -- $10^{-5}$
(Heckman et al. 1982; van Breugel et al. 1985).
While the same value is observed in a high-redshift radio galaxy (Rush et
al.\ 1997), we adopted the value $f=10^{-4}$.
However, the reader should keep in mind that this value is
uncertain and may differ by an order of magnitude.

The mass of the line emission gas was calculated using the relation
\begin{equation}
M_{\rm gas}=V f n_{\rm e} m_{\rm H},
\end{equation}
where $m_{\rm H}$ is the hydrogen mass.
We next assumed that the line emission gas was filling the cones at the
beginning, and was swept up by the passage of radio jets. 
We thus deduced the total mass of the gas surrounding the galaxy
$\displaystyle M_{\rm tot}=M_{\rm gas}\frac{4\pi}{\Omega}$, taking the
opening angles of the 
cones $\Omega$ to be 0.12$\pi$ str and 0.06$\pi$ str for the northern and
southern cones, respectively. 

The number of ionizing photons $Q$ can be written as (Osterbrock 1989) 
\begin{equation}
 Q=\frac{\alpha_{\rm B}}{\alpha_{\rm H\alpha}^{\rm eff}}\frac{L({\rm H}\alpha)}{h \nu_{\rm H \alpha} },
\label{eq_photon}
\end{equation}
where $\alpha_{\rm B}=1.43\times10^{-13}$ is the total recombination
coefficient under the case-B condition. 
The total number of ionizing photons radiated by the central engine
was calculated as $\displaystyle Q_{\rm tot}=Q\frac{4\pi}{\Omega f_{\rm c}}$,
where $f_{\rm c}$ is the covering factor of the cloud in the cone.

At last, the ionization parameter of the clouds is defined as
\begin{equation}
 U=\frac{Q}{R^2 \Omega f_{\rm c} n_{\rm e} c},
\end{equation}
where $R$ is the distance of the cloud from the central engine and
$c$ the velocity of light.
Because we do not know the covering factor, $f_{\rm c}$, we calculated the
lower limits of $Q_{\rm tot}$ and $U$ assuming $f_{\rm c}=1$. 

All of these calculated values are given in table 4.
The electron density is on the order of 50 (cm$^{-3}$), the ionization
parameter 0.001, and the mass of the emission line cloud
$10^{9}~M_{\odot}$. These values are similar to those deduced for
the $\sim$100 kpc extended line emission clouds of other high-redshift
radio galaxies (Rush et al.\ 1997; Heckman et al.\ 1991). 
The large luminosity of the ionizing photons ($>10^{57}$ photons
$\rm s^{-1}$) draws our attention, which is far larger than the values
for typical low-$z$ radio galaxies ($10^{51-54}$ photons $\rm s^{-1}$)
(Baum, Heckman 1989), even larger than that of 
radio loud QSOs ($10^{54-56}$ photons $\rm s^{-1}$)
(Stockton, MacKenty 1987), and almost comparable to that of the brightest
QSOs ($10^{57-58}$ photons $\rm s^{-1}$) (Hill et al.\ 1993).
Here, we calculated the luminosity of ionizing photons for QSOs from
their $L(\rm H\alpha)$ using equation (\ref{eq_photon}), or from $L(\rm
H\beta)$ with $L({\rm H}\alpha)/L({\rm H}\beta)=4 $ taken from
Kwan and Krolik (1981). 

Next, we carried out a photoionization calculation using the code
CLOUDY90 (Ferland et al.\ 1998) by assuming a wide range of hydrogen
density, $n_{\rm H}$, and ionization parameter, $U$.
The cloud distance from an ionizing source was set to 25 kpc and
its thickness to 100 pc. 
We assumed the metal abundance to be $Z=0.1 Z_{\odot}$, according to the
observations of $z\sim2$ damped Ly$\alpha$ clouds (Pettini et al. 1994).
The continuum spectrum of the ionizing source is set to a power law,
$f_{\nu}\propto\nu^{\alpha}$, with $\alpha=-0.7$ longward of $912~\rm \AA$ and $\alpha=-2.5$ shortward of  $912~\rm \AA$, taken from the composite
spectrum of radio-loud QSOs (Cristiani, Rio 1990; Zheng et al.\ 1997). 
The resultant equivalent widths, calculated from the emission-line
strength and the diffuse continuum radiation, are shown in table 5.
Most of the simulated equivalent widths exceed 1000 $\rm \AA$, satisfying the
observed value.

Thus, we suggest that the emission mechanism of the extended
H$\alpha$+[N {\sc ii}] cones of B3~0731+438 is nebular emission from
clouds ionized by the strong UV radiation of the hidden AGN.

\subsection{Spectral Energy Distribution}

Assuming that the exponential disk component of the $K$-continuum image is a 
host galaxy of B3~0731+438 and the compact core a type-2 AGN,
we reconstructed the SED of the B3~0731+438 with model spectra of a
type-2 AGN and a galaxy. 
Because contamination of the [O {\sc ii}] $\lambda\lambda$ 3727 emission
line to the $J$-band flux is expected, we subtracted it,
assuming an H$\alpha$/[O {\sc ii}] ratio of 3 (McCarthy et al.\ 1995).
The model spectrum of the galaxy was calculated by the spectrophotometric 
galaxy evolution model PEGASE (Fioc, Rocca-Volmerange 1997)
under three variations of star-forming history, namely,
an instantaneous burst model, and two exponential burst models
with time scales of $\tau=200$ Myr and 2 Gyr, respectively.
For the model spectrum of the type-2 AGN, we selected the dust-scattered
AGN model calculated by Cimatti et al.\ (1994) with a scattering
angle, $\Theta$, of $90^{\circ}$ and the power-law index of the incident
continuum being $\alpha=-0.7$, and assuming extinction of SMC dust
(Pr\'evot et al.\ 1984, Bouchet et al.\ 1985). 
We scaled these two spectra according to their observed $K$-continuum flux and
fit the $J$- and $R$-band flux by altering the age of the
galaxy and the extinction of the type-2 AGN. 

We show the results in figure 4 and table 6.
The $\tau=$ 2 Gyr exponential burst model is not plausible, because the
age of the best-fit galaxy model is 10 Gyr, which is larger than the
cosmic age (3.5 Gyr for the assumed cosmological parameters).
Both the $\tau=$ 200 Myr exponential burst model and the instantaneous
burst model fit the SED. 
However, we prefer the instantaneous burst model because the observed
flux of the $R$-band appears to be dominated by the diffuse galactic
component, as can be seen in the $R$-band image of McCarthy (1991).

Consequently, we suppose that the age of the galaxy is 500 Myr,
resulting in a formation redshift of $z_{\rm form}\sim3$.
The total stellar mass of the galaxy estimated from the model is 
3$\times10^{11}~M_{\odot}$, comparable to that of a typical
3C radio galaxy at $z=1$ (Best et al.\ 1997).

The H$\alpha$+[N {\sc ii}] peak is located at the center of the galaxy
and aligned to the radio axis.
Infrared spectroscopic observations showed that the H$\alpha$ line width
is narrow ($ < 560\rm~ km~s^{-1}$ ; McCarthy et al. 1992) and 
that the emission-line ratios are similar to those of Seyfert 2 galaxies
(Evans 1997). 
We therefore suggest that the H$\alpha$+[N {\sc ii}] luminosity at the
center of the
galaxy is dominated by scattered radiation from the narrow-line regions 
of the hidden nucleus.
However, the possibility of intense starburst activity cannot be ruled
out entirely, and polarimetry is necessary for a confirmation.

\section{Summary and Conclusions}
We observed the powerful radio galaxy B3~0731+438 in the infrared
$K^{\prime}$-band and the 2.25 $\mu$m narrow-band, corresponding to the
rest-frame 6000 $\rm \AA$ 
continuum and H$\alpha$+[N {\sc ii}] emission line, respectively.
We then produced a line-free $K$-continuum image and a
continuum-subtracted H$\alpha$+[N {\sc ii}] emission-line image.

Our observations are the first to show a cone-shaped H$\alpha$+[N {\sc ii}]
emission-line structure of a high-redshift radio galaxy at infrared
wavelength. The radio-aligned cones suggest that the gas is ionized by
the hidden AGN. 
The contribution of the scattered light from the hidden AGN
to the H$\alpha$ luminosity is estimated to be small.
On the other hand, we find that a gas cloud ionized by a power-law UV
continuum can account for the large observed H$\alpha$+[N {\sc ii}] equivalent
width, using the values of the electron density and the ionization parameter
estimated from the observed H$\alpha$ line luminosity.
Together with their biconical structure, we infer that we have detected
H$\alpha$ ionization cones of a high-redshift powerful radio galaxy
for the first time. 
The estimated mass of the ionized gas cones is of the order of $10^{9}~M_{\odot}$,
and the expected total mass of the gas surrounding the galaxy is
$10^{11}~M_{\odot}$.

The H$\alpha$+[N {\sc ii}] peak at the center of the galaxy is also aligned to the
radio axis. We suppose that this peak is scattered light from the narrow-line
regions of the hidden AGN, and that polarimetry is necessary for a
confirmation.

The $K$-continuum image is separated into two components, 
assumed to be a type-2 AGN and an underlying host galaxy.
The SED of the whole radio galaxy is modeled by a model spectrum of a 500
Myr-old instantaneous burst galaxy and dust-scattered
power-law continuum with $A_V=1.9$ mag extinction. 
The stellar mass of the galaxy is $3\times10^{11}~M_{\odot}$, which is
comparable to that of a typical 3C radio galaxy at $z=1$.

\vspace{1pc}\par
We thank all staff of the Subaru Telescope, who supported us to set
up our instrument, and helped us to make these observations.
We would like to express our thanks to the engineering staff of Mitsubishi 
Electric Co. for their fine operation of the telescope during the test
observation runs, and the staff
of Fujitsu Co. for timely provision of control software.

We also appreciate  M. Fioc and B. Rocca-Volmerange for generously
 offering their galaxy modeling code, PEGASE, and
G. Ferland for the spectral synthesis code, CLOUDY90.

K.Motohara was financially supported by the Japan Society for the
Promotion of Science.
\clearpage
\section*{References} 
\re Armus L., Soifer B.T., Murphy T.W.Jr, Neugebauer G., Evans A.S.,
Matthews K.\ 1998, ApJ 495, 276
\re Baker J.C., Hunstead R.W., Kapahi V.K., Subrahmanya C.R.\ 1999,
ApJS 122, 29
\re Baum S.A., Heckman T.\ 1989, ApJ 336, 681
\re Begelman M.C., Cioffi D.F.\ 1989, ApJ 345, L21
\re Best P.N., Longair M.S., R\"ottgering H.J.A.\ 1997, MNRAS 292, 758
\re Binette L., Magris G., Bruzual G.\ 1992, in Relationship Between
Active Galactic Nuclei and Starburst Galaxies, ed A. V. Filippenko, ASP
Conf. Ser. 31, 211
\re Bouchet P., Lequeux J., Maurice E., Pr\'evot L., Pr\'evot-Burnichon
M.L.\ 1985, A\&A 149, 330
\re Calzetti D.\ 1997, AJ 113, 162
\re Carilli C.L., R\"ottgering H.J.A., van Ojik R., Miley G.K., van
Breugel W.J.M.\ 1997, ApJS 109, 1
\re Chambers K., Miley G., van Breugel W.\ 1987, Nature 329, 604
\re Charlot S., Fall S.M.\ 1993 ApJ 415, 580
\re Cimatti A., di Serego Alighieri S., Field G.B., Fosbury R.A.E.\ 1994,
			     ApJ 422, 562
\re Cristiani S., Vio R.\ 1990, A\&A 227, 385
\re Daly R.A.\ 1992a, ApJ 386, L9
\re Daly R.A.\ 1992b, ApJ 399, 426
\re de Young D.S.\ 1989, ApJ 342, L59
\re Dickson R., Tadhunter C., Shaw M., Clark N., Morganti R.\ 1995, MNRAS
			      273, L29 
\re di Serego Alighieri S., Fosbury R.A.E., Quinn P.J., Tadhunter C.N.\ 1989,
Nature 341, 307
\re Eales S.A., Rawlings S.\ 1993, ApJ 411, 67
\re Egami E., Armus L., Neugebauer G., Soifer B.T., Evans A.S., Murphy
 T.W.Jr\ 1999, in The Hy-Redshift Universe: Galaxy Formation
 and Evolution at High Redshift, ed A.J. Bunker and W.J.M. van Breugel, 
 ASP Conf. Ser. in press (astro-ph/9909517)
\re Espey B.R., Carswell R.F., Bailey J.A., Smith M.G., Ward M.J.\
			      1989, ApJ 342, 666
\re Evans A.S.\ 1998, ApJ 498, 553
\re Fabian A.C.\ 1989, MNRAS 238, 41p
\re Ferland G.J., Korista K.T., Verner D.A., Ferguson J.W., Kingdon J.B.,
			     Verner E.M.\ 1998, PASP 110, 761
\re Fioc M., Rocca-Volmerange B.\ 1997, A\&A 326, 950
\re Heckman T.M., Lehnert M.D., Miley G.K., van Breugel W.\ 1991, ApJ
381, 373
\re Heckman T.M., Miley G.K., Balick B., van Breugel W.J.M., Butcher
H.R.\ 1982, ApJ 262, 529
\re Hill G.J., Thompson K.L., Elston R.\ 1993, ApJ 414, L1
\re Kwan J., Krolik J.H.\ 1981, ApJ 250, L478
\re McCarthy P.J.\ 1991 AJ 102, 518
\re McCarthy P.J., Elston R., Eisenhardt P.\ 1992, ApJ 387, L29
\re McCarthy P.J., Spinrad H., van Breugel W.\ 1995 ApJS 99, 27
\re McCarthy P.J., van Breugel W., Spinrad H., Djorgovski S.\ 1987,
			     ApJ 321, L29
\re McQuade K., Calzettei D., Kinney A.L.\ 1995, ApJ 97, 331
\re Motohara K., Maihara T., Iwamuro F., Oya S., Imanishi M., Terada H.,
Goto M., Iwai J. et al.\ 1998, Proc.\ SPIE 3354, 659 
\re Osterbrock D.E.\ 1989,  Astrophysics of Gaseous Nebulae and Actice Galactic
			      Nuclei (University Science Books, Mill
			      Valley)
\re Pettini M., Smith L.J., Hunstead R.W., King D.L.\ 1994, ApJ 426, 79
\re Pr\'evot M.L., Lequeux J., Maurice E., Pr\'evot L., Rocca-Volmerange
			      B.\ 1984, A\&A 132, 389  
\re Rees M.J.\ 1989, MNRAS 239, 1p
\re Rush B., McCarthy P.J., Athreya R.M., Persson S.E.\ 1997, ApJ 484, 163
\re Stockton A., MacKenty J.W.\ 1987, ApJ 316, 584
\re van Breugel W., Miley G., Heckman T., Butcher H., Bridle A.\ 1985,
ApJ 290, 496
\re Zheng W., Kriss G.A., Telfer R.C., Grimes J.P., Davidsen A.F.\ 1997, ApJ
			      475, 469

\clearpage

\begin{table*}[t]
\begin{center}
Table~1.\hspace{4pt}Observational logs of B3~0731+438.\\
\vspace{6pt}
\begin{tabular}{ccccc}
\hline\hline
Date & Object & Band & Exposure  & Seeing \\
\hline
1999 February 25 &B3~0731+438 &N225 & 60 s $\times$ 48 & 0$.\!\!^{\prime\prime}$6 \\
& FS 21 & N225 & 10 s $\times$ 24 &\\
1999 February 27 & FS 15 & $K^{\prime}$ & 2 s $\times$ 24 &\\
& FS 15 &N225 & 10 s $\times$ 24 &\\
&B3~0731+438 & $K^{\prime}$ & 20 s $\times$ 96 & 0$.\!\!^{\prime\prime}$4 \\
&B3~0731+438 & N225 & 60 s $\times$ 48 & 0$.\!\!^{\prime\prime}$4 \\
& FS 23 & $K^{\prime}$ & 10 s $\times$ 24 &\\
\hline
\end{tabular} \end{center}
\end{table*}

\begin{table*}[t]

\begin{center}
Table~2.\hspace{4pt}Photometric data of B3~0731+438.\\
\vspace{6pt}
{
\begin{tabular}{p{7pc}cccccc}
\hline\hline
\multicolumn{1}{c}{Band} & $\lambda$ & $\Delta\lambda$ &
 $F_{\nu}$ & Mag & Aperture radius  &
 References\\
& ($\mu$m) & ($\mu$m) & ($\mu$Jy) && ($^{\prime\prime}$) &\\
\hline
$R$\dotfill & 0.70 &0.22& $1.1 \pm 0.3$ &23.6& 1$.\!\!^{\prime\prime}$5 $\times$ 3$.\!\!^{\prime\prime}$5$^{*}$ & McCarthy (1991)\\
$J$\dotfill & 1.25 &0.38& $15.0 \pm 3.8$ &20.1& 4.7 & Iwamuro, private comm.\\
$K^{\prime}$\dotfill & 2.13 &0.34& $43.7 \pm 2.3$ &18.0&4.6 & \\
N225\dotfill & 2.25 &0.022& $302 \pm 7.4 $  &--& 4.6 \\
\hline
$K$-continuum core\dotfill &2.13& 0.32 & $12.6 \pm 2.5 $ &19.3& 4.7 &\\
$K$-continuum diffuse\dotfill &2.13& 0.32 & $15.0 \pm 2.5 $ &19.1& 4.7 &\\
\hline
\end{tabular}}
 \end{center}
\vspace{6pt}\par\noindent
$*$  Flux density is extracted from the spectrophotometric data of
 1$.\!\!^{\prime\prime}$5 slit width and 3$.\!\!^{\prime\prime}$5 spatial aperture. \\
\label{allphotometry}
\end{table*}

\begin{table*}[t]
\begin{center}
Table~3.\hspace{4pt}Emission-line properties of B3~0731+438.\\
\vspace{6pt}
\begin{tabular}{p{6pc}cccc}
\hline\hline
\multicolumn{1}{c}{Area} & Size$^{*}$ & $F_{\lambda}(\rm continuum \it) \rm^{\dagger}$   & $F(\rm line \it) \rm^{\ddagger}$ &  $W(\rm line)\it \rm^{\S}$\\
& ($^{\prime\prime}\times^{\prime\prime}$) & {$(\rm W~m^{-2}\it \mu\rm m^{-1}) $} & {$(\rm W~m^{-2})$} & {$(\rm \AA)$}\\
\hline
Northern Cone\dotfill &$3.596\times1.740$& $8.45(\pm 4.08)\times 10^{-19}$ & $ 3.18(\pm 0.26)\times 10^{-19}$ &$1097^{+1199}_{-417}$\\
Southern Cone\dotfill &$1.856\times2.204$& $10.0(\pm 3.29)\times 10^{-19}$ & $ 3.80(\pm 0.21)\times 10^{-19}$ &  $1098^{+645}_{-310}$\\
Central Core\dotfill&$2.436\times2.436$& $1.47 (\pm 0.03)\times 10^{-17}$ & $ 1.74(\pm 0.03)\times 10^{-18}$ & 344$\pm$14 \\
\hline
\end{tabular} \end{center}
\vspace{6pt}\par\noindent $*$~ Size of the aperture used for photometry.
\par\noindent
 $\dagger$~ Flux density of the continuum emission.
\par\noindent
 $\ddagger$~ Flux of the H$\alpha$+[N {\sc ii}] emission lines.
\par\noindent
 $\S$~ Rest-frame equivalent width of the H$\alpha$+[N {\sc ii}] emission lines.
\end{table*}
\clearpage
\begin{table*}[t]
\begin{center}
Table~4.\hspace{4pt}Physical properties in the emission-line cones of
 B3~0731+438 assuming a filling factor of $10^{-4}$.\\
\vspace{6pt}
\begin{tabular}{p{4pc}cccccccc}
\hline\hline
\multicolumn{1}{c}{Area} &$V$ & $n_{\rm e}$  & $M_{\rm gas}$ & $M_{\rm tot}$&
 $Q$ &$Q_{\rm tot}$ & $U$\\
 & (cm$^3$) & ($\rm cm^{-3}$) & ($M_{\odot}$) & ($M_{\odot}$) & (photons 
 s$\rm ^{-1}$)& (photons s$\rm ^{-1}$)\\
\hline
North\dotfill &$8.5\times10^{68}$& $3.8\times 10$ & $2.8\times 10^{9}$ &
 $6.4\times 10^{10}$ &  $1.8\times 10^{55}$&  $>6.1\times 10^{56}$ &
 $>1.3\times10^{-3}$\\
South\dotfill &$5.1\times10^{68}$& $6.8\times 10$ & $1.9\times 10^{9}$ &
 $8.9\times 10^{10}$ &  $2.0\times 10^{55}$ &  $>1.4\times 10^{57}$ & $>5.9\times10^{-3}$\\
\hline
\end{tabular} \end{center}
\end{table*}

\begin{table*}[t]
\begin{center}
Table~5.\hspace{4pt}Simulated equivalent widths (in $\rm \AA$) of
 H$\alpha$+[N {\sc ii}] emission line by  CLOUDY90.\\
\vspace{6pt}
\begin{tabular}{|c|c|ccc|}
\cline{3-5}
\multicolumn{2}{c|}{}&\multicolumn{3}{c|}{$n_{\rm H}$(cm$^{-3}$)}\\
\cline{3-5}
\multicolumn{2}{c|}{}& 100 & 10 & 1 \\
\hline
 &1      &762  &822  &1054 \\
 &0.1    &1229 &973  &952  \\
$U$&0.01 &\underline{1301} &\underline{1300} &1119 \\
 &0.001  &\underline{1562} &\underline{1561} &1561  \\
 &0.0001 &1842 &1842 &1842 \\
\hline
\end{tabular} \end{center}
\vspace{6pt}\par\noindent $*$~ 
 Underlined are those of
 estimated physical parameters. 
\end{table*}
\begin{table*}[t]
\begin{center}
Table~6.\hspace{4pt}Parameters for the best-fit models of the spectral
 energy distribution. \\
\vspace{6pt}
\begin{tabular}{p{12pc}ccc}
\hline
\multicolumn{1}{c}{Model} & $A_{V}$ $^*$ & Age (Myr)$^{\dagger}$ & Stellar Mass ($M_{\odot}$)$^{\ddagger}$ \\
\hline
Instantaneous\dotfill & 1.9 & 500 & $3.4\times10^{11}$ \\
Exponential ($\tau=200$ Myr)\dotfill & 1.4 & 2000 & $6.5\times10^{11}$\\
Exponential ($\tau=2$ Gyr)\dotfill & 1.4 & 10000 & $1.7\times10^{12}$ \\
\hline
\end{tabular} \end{center}
\vspace{6pt}\par\noindent
 *~ Extinction of the type 2 AGN.
\par\noindent
 $\dagger$~ Age of the galaxy.
\par\noindent
 $\ddagger$~ Stellar mass of the galaxy.
\end{table*}
\clearpage
\onecolumn
\bigskip

Figure 1.~~Color image of B3~0731+438,
 11$.\!\!^{\prime\prime}$8 on a side. The blue color is assigned to the
 line-subtracted $K^{\prime}$-band image and the red color to the
 2.25 $\mu$m narrow-band. Both images were smoothed by a 1 pixel Gaussian
 filter. Green is assigned to the average of the red and the blue
 image. 
\\\\
\bigskip
\epsfxsize=0.8\textwidth
\centerline{\epsfbox{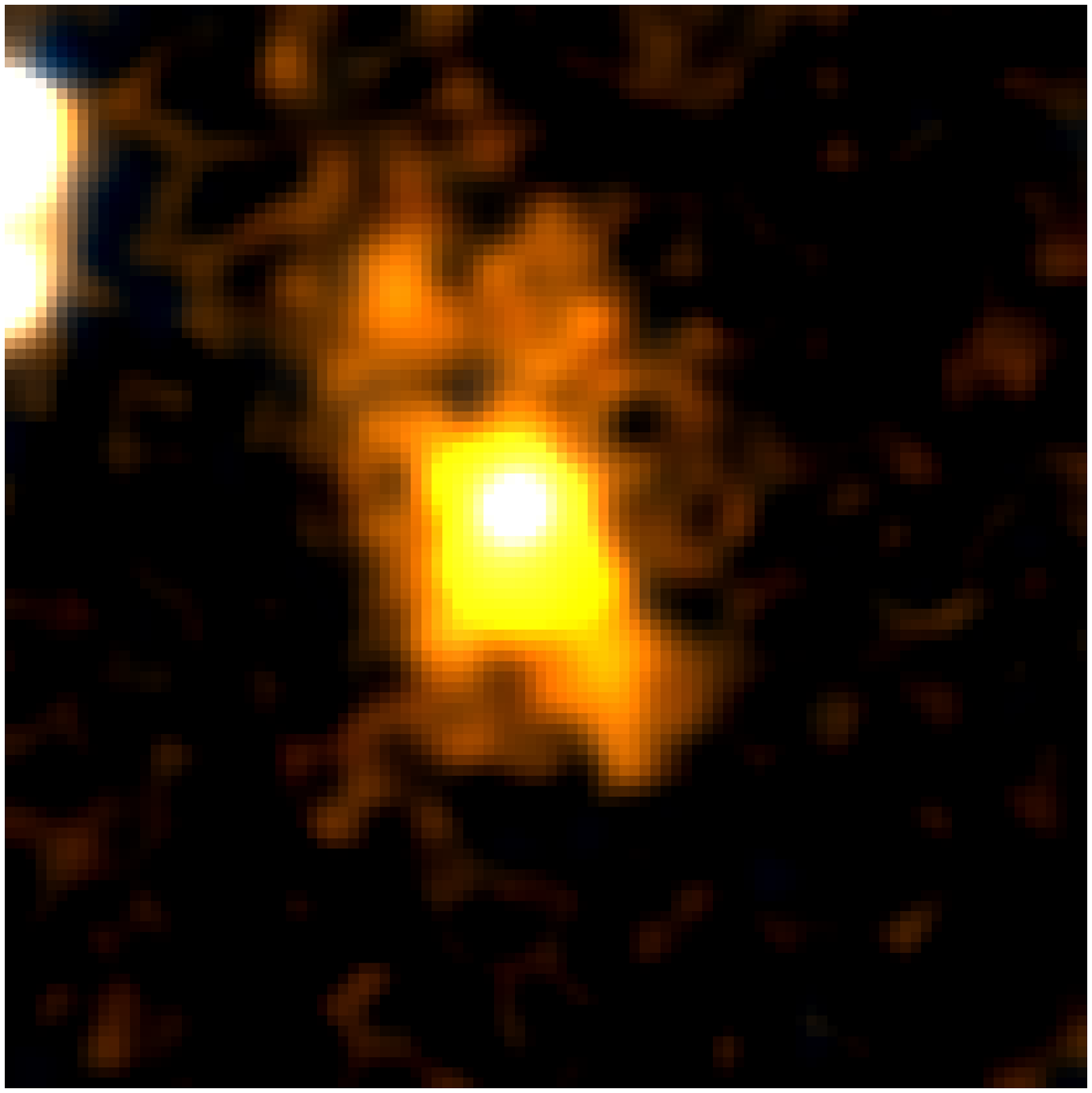}}

\clearpage
\bigskip
Figure 2.~~Logarithmic gray-scale images of B3~0731+438. The squares are the
 apertures with which aperture photometry was carried out,
 while the crosses indicate the
 position of 4710 MHz radio hot spots, as mapped by Carilli et
 al. (1997).
(a) The H$\alpha$+[N {\sc ii}] image, smoothed by a Gaussian filter of 1 pixel. 
 The first contour is at
 1 $\sigma$ above the background level, while the subsequent contours
 are at levels of $2n\sigma~(n=1,2,3,...)$. $\sigma$ is taken from the
 non-smoothed image. 
(b) The $K$-continuum image, which was smoothed by a 3 pixel Gaussian
 filter to match
 the seeing size with the H$\alpha$+[N {\sc ii}] image.
 The contours are at levels of $n\sigma~(n=1,2,3,...)$, where $\sigma$ is
 again taken from the non-smoothed image.
 The profile at the lower-right corner is the contours of the stellar
 image used for deconvolution, whose peak value is normalized to B3
 0731+438. (c) The $K$-continuum image deconvolved by MEM
 deconvolution method.
\\\\
\bigskip
\epsfxsize=0.9\textwidth
\centerline{\epsfbox{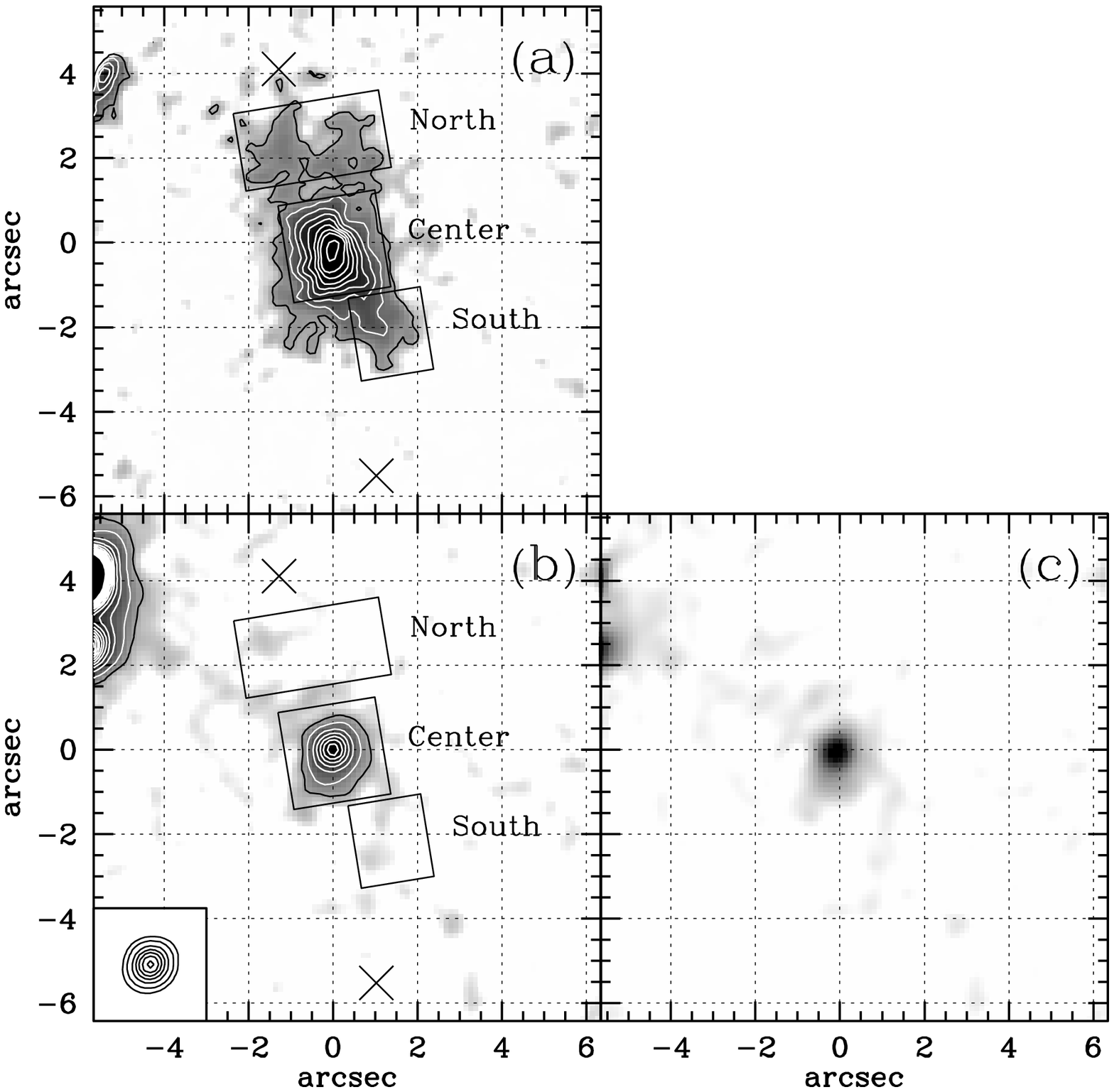}}

\clearpage
\bigskip
Figure 3.~~(upper-left) The $K$-continuum image of B3
 0731+438. (upper-right) The best-fit two-component model profile of
 the $K$-continuum image.
(lower-left) The stellar-profile subtracted $K$-continuum image, which
 reveals  the underlying galaxy. 
(lower-right) The residual image after the two-component profile is
 subtracted. \\\\
\bigskip
\epsfxsize=0.9\textwidth
\centerline{\epsfbox{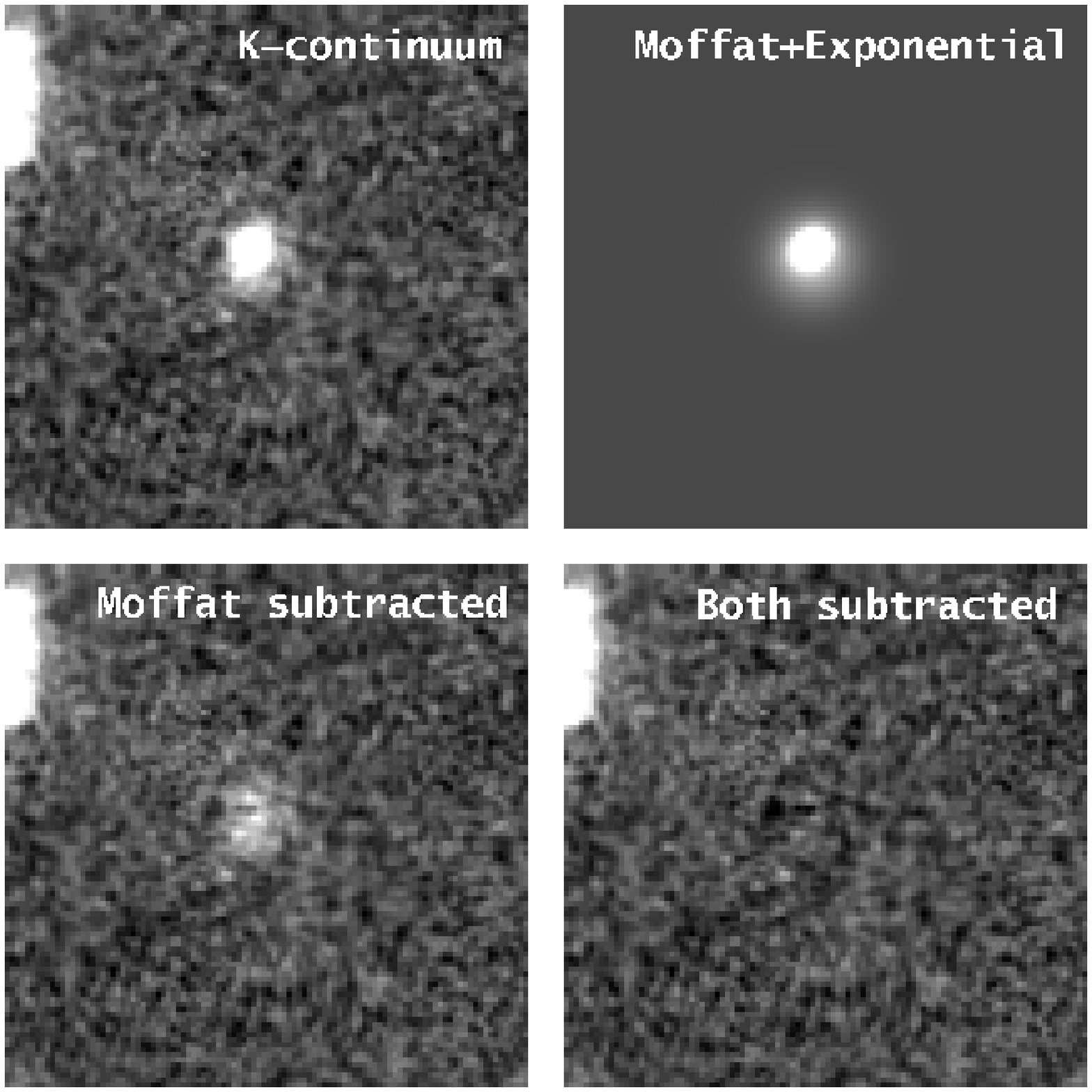}}
\clearpage

\bigskip
Figure 4.~~Spectral energy distribution of B3~0731+438. The open circle is
 from McCarthy (1991), the open square from Iwamuro (private
 communication),  and the filled square from this work. The thick solid
 line is the best-fit models, and the thin solid line the spectra of (A)
 the type-2 AGN with  dust extinction and (B) the galaxy-evolution model. 
 The dotted line is the spectrum of the type-2 AGN without extinction. 
 The star-formation models are: (a) instantaneous burst and (b) $\tau=200$
 Myr exponential  burst.
\\\\
\bigskip
\epsfxsize=0.9\textwidth
\centerline{\epsfbox{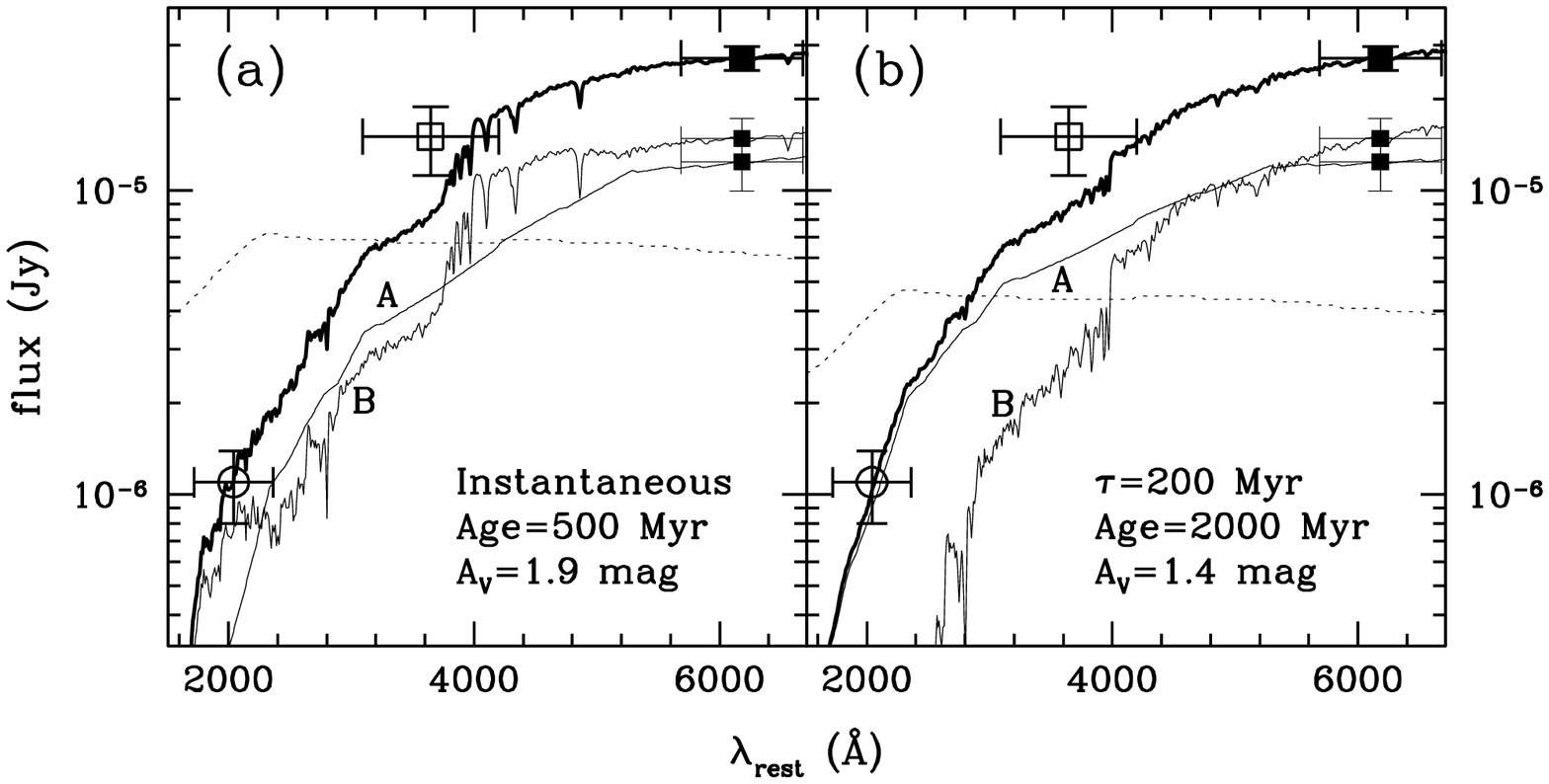}}
\end{document}